\documentclass{aastex}
\usepackage{emulateapj5}
\usepackage{apjfonts}
\usepackage{epsfig}

\slugcomment{Received 2006 May 15; accepted 2006 June 28 }
\journalinfo{The Astrophysical Journal Letters in Press}

\shorttitle{Duty Cycle of Quasars}
\shortauthors{Wang, Chen \& Zhang}

\def\calcb{{\cal C}_{\rm B}}
\def\caln{{\cal N}(\mbh,z)}

\def\kms{\ifmmode {\rm km~ s^{-1}} \else {\rm km~s^{-1}}\ \fi}
\def\dotmbh{\dot{M}_{\bullet}}
\def\mbh{M_{\bullet}}
\def\mmbh{m_{\bullet}}

\def\mgii{\ifmmode Mg {\sc ii} \else Mg {\sc ii}\ \fi}
\def\oiii{\ifmmode O {\sc iii} \else O {\sc iii}\ \fi}
\def\feii{\ifmmode Fe {\sc ii} \else Fe {\sc ii}\ \fi}

\def\rmd{{\rm d}}

\def\sunm{M_{\odot}}

\def\tp{t^{\prime}}
\def\tpi{t_i^{\prime}}

\def\lax{{$\mathrel{\hbox{\rlap{\hbox{\lower4pt\hbox{$\sim$}}}\hbox{$<$}}}$}}
\def\gax{{$\mathrel{\hbox{\rlap{\hbox{\lower4pt\hbox{$\sim$}}}\hbox{$>$}}}$}}

\begin{document}

\title{Cosmological Evolution of the Duty Cycle of Quasars}

\author{Jian-Min Wang\altaffilmark{1}, Yan-Mei Chen\altaffilmark{1,2} and Fan Zhang\altaffilmark{1}}

\altaffiltext{1}{Key Laboratory for Particle Astrophysics, Institute of High Energy Physics, 
                 Chinese Academy of Sciences, 19B Yuquan Road, Beijing 100049, China}

\altaffiltext{2}{Graduate School, Chinese Academy of Science, 19A Yuquan Road, Beijing 100049, China}

\begin{abstract}
Quasars are powered by accretion onto supermassive black holes, 
but the problem of the duty cycle related to the episodic activity of the black holes remains open as one of 
the major questions of cosmological evolution of quasars. In this Letter, we obtain quasar
duty cycles based on analyses of a large sample composed of 10,979 quasars with redshifts $z\le2.1$ 
from the Sloan Digital Sky Survey (SDSS) Data Release Three. We estimate masses of quasar black holes 
and obtain their mass function (MF) of the present sample.  We then get the duty cycle 
$\bar{\delta}(z)=10^{-3}\sim 1$ based on the So\l tan's argument, implying that black holes are undergoing 
multiple episodic activity. We find that the duty cycle has a strong evolution.
By comparison, we show that evolution of the duty cycle follows the history of cosmic star 
formation rate (SFR) density in the Universe, providing intriguing evidence for a natural 
connection between star formation and triggering of black hole activity.
Feedback on star formation from black hole activity is briefly discussed.
\end{abstract}
 
\section{Introduction}
Supermassive black holes are relics of quasars  in the Universe (So\l tan 1982; Rees 1984, 1990). 
Evolution of quasars is led by switching on and off accretion onto the black holes. 
During their entire evolution, how many times and how many black holes are triggered? what is
the triggering mechanism? and why do quasars switch off?

The duty cycle, defined as the fraction of active black holes to their total number
is a key to tackling the above problems (Richstone et al. 1998; Martini 2004). 
A popular method to get the duty cycle invokes the continuity equation and the MF of quasar
black holes. With an assumption that quasar black holes have the same Eddington ratio,
their MF can be obtained from the luminosity function and finally the duty cycle can be found
from the continuity equation (Small \& Blandford 1990; Marconi et al. 2004). 
This is a convenient way to discuss evolution of the black holes, but 
the degeneracy of the Eddington ratio and the duty cycle still holds.
Actually, not only are the Eddington ratios {\em not} constant for different quasars 
at different epochs, but they appear to be quite scattered (Vestergaard 2004). 
The duty cycle is poorly understood as a statistical parameter tracing the evolution of quasar 
populations.

In recent years, there has been much progress in estimating black hole masses both in nearby 
galaxies and distant quasars. The empirical relation of reverberation mapping allows us to conveniently 
obtain the black hole masses from a large sample and directly get their MF. Thus it becomes 
realistic to get new clues to understand the evolution of quasars. Fortunately,  
by invoking the MF, we can decouple the degeneracy of the duty cycle and
the Eddington ratio to get the duty cycle.

In this Letter, we use available SDSS data to directly get the MF of 
the black holes so as to discuss the duty cycle based on the MF and find
that there is a strong cosmological evolution. Our calculations assume a cosmology with the
Hubble constant $H_0=70~{\rm Mpc^{-1}~km~s^{-1}}$, $\Omega_{\rm m}=0.3$, and $\Omega_{\Lambda}=0.7$.

\section{Black Hole Evolution}
Active black hole evolution can be described by the MF $\Phi(\mbh,z)$, which is defined as
$\Phi(\mbh,z)=\rmd^2 N/\rmd\mbh\rmd V,$ where $\mbh$ is the black hole 
mass, $\rmd N$ is the number of quasars within the comoving volume element 
$\rmd V$ and mass interval $\rmd \mbh$. The number density of quasar black holes is then given by
$N_{\rm qso}(z)=\int_{\mbh^*}\Phi(\mbh,z)\rmd \mbh$ at redshift $z$, where $\mbh^*$ is the flux-limited
black hole mass in a survey. If $\caln$ is the MF of 
all black holes including the active and the inactive ones, the total number density of black holes is
$N_{\rm tot}(z)=\int_{\mbh^*}\caln\rmd\mbh$ at redshift $z$. The relation between $\caln$ and 
$\Phi(\mbh,z)$ can be written as (Small \& Blandford 1992; Marconi et al. 2004),  
\begin{equation}
\Phi(\mbh,z)=\delta(\mbh,z)\caln. 
\end{equation}
where the duty cycle, $\delta(\mbh,z)$ is a function of black hole mass and redshift. Integrating 
equation (1) over all black hole masses, we have a mean duty cycle in terms of their number density
\begin{equation}
{\bar{\delta}}_1(z)=\frac{\int_{\mbh^*}\Phi(\mbh,z)\rmd \mbh}{\int_{\mbh^*}\caln\rmd \mbh}=
\frac{N_{\rm qso}(z)}{N_{\rm tot}(z)}.
\end{equation}
This duty cycle represents the relative number of the active black holes
to the total. Multiplying by $\mbh$ and integrating Eq. (1) over all black hole masses, we have an
averaged duty cycle weighted by the masses of the black holes
\begin{equation}
{\bar \delta}_2(z)=\frac{\int_{\mbh^*}\Phi(\mbh,z)\mbh\rmd\mbh}{\int_{\mbh^*}\caln\mbh\rmd \mbh}
=\frac{\langle\mbh(z)\rangle_{\rm qso}}{\langle\mbh(z)\rangle_{\rm all}}
{\bar{\delta}}_1(z),
\end{equation}
where $\langle{\mbh(z)}\rangle_{\rm qso}=\int_{\mbh^*}\Phi(\mbh,z)\mbh\rmd \mbh/N_{\rm qso}$ 
is the averaged mass of the quasar black holes at redshift $z$ and
$\langle{\mbh(z)}\rangle_{\rm all}=\int_{\mbh^*}\caln\mbh\rmd \mbh/N_{\rm tot}$ for all of active 
and inactive black holes. We show below $\langle{\mbh(z)}\rangle_{\rm qso}=\langle{\mbh(z)}\rangle_{\rm all}$
for $\mbh>\mbh^*$.
So we have $\bar{\delta}_1(z)=\bar{\delta}_2(z)$.

The averaged mass of all black holes is 
$\langle\mbh(t)\rangle_{\rm all}=\langle\mbh^{\rm S}\rangle+\int_0^t\langle\dotmbh(\tp)\rangle_{\rm all} 
\rmd t^{\prime}$, where $\langle\mbh^{\rm S}\rangle$ is the mean mass of seed black holes and
$\langle\dotmbh(\tp)\rangle_{\rm all}$ is the averaged accretion rate of all
black holes larger than $\mbh^*$. Considering that 
$\langle\dotmbh(\tp)\rangle_{\rm all}=\bar{\delta}_1(\tp)\langle\dotmbh(\tp)\rangle_{\rm qso}$ 
[see Eq. (9) in Small \& Blandford 1992; and the first equation in sentence with Eq. (14) in 
Marconi et al. 2004], we have  
$\langle\mbh(t)\rangle_{\rm all}=\langle\mbh^{\rm S}\rangle+
\int_0^t\bar{\delta}_1(\tp)\langle\dotmbh(\tp)\rangle_{\rm qso} \rmd \tp$. On the other hand,
the duty cycle can be written as $\bar{\delta}_1(t)=\Delta t_{\rm act}/\Delta t$
for a single episodic phase within a time interval $\Delta t$, where 
$\Delta t=\Delta t_{\rm act}+\Delta t_{\rm dor}$; $\Delta t_{\rm act}$ and $\Delta t_{\rm dor}$ are
the active and the dormant times, respectively. Then, for the $i-$th episodic phase within $\Delta \tpi$, 
the accreted mass is given by $\langle\dotmbh(\tpi)\rangle_{\rm qso}\bar{\delta}_1(\tpi)\Delta \tpi$. The 
mean mass of quasar black holes
then reads $\langle\mbh(t)\rangle_{\rm qso}=\langle\mbh^{\rm S}\rangle+
\sum_i \langle\dotmbh(\tpi)\rangle_{\rm qso} \bar{\delta}_1(\tpi)\Delta t_i^{\prime}
=\langle\mbh^{\rm S}\rangle+\int_0^t\langle\dotmbh(\tp)\rangle_{\rm qso} 
\bar{\delta}_1(\tp)\rmd t^{\prime}$ after multiple episodic phases. We then have
$\langle\mbh(t)\rangle_{\rm all}=\langle\mbh(t)\rangle_{\rm qso}$.
This relation can be understood more easily if we consider multiple episodic growth of black holes. 
We thus have the duty cycle of the black holes in a concise form of
$\bar{\delta}(z)\equiv \bar{\delta}_1(z)=\bar{\delta}_2(z)$.


The total mass density of the black holes at redshift $z$ contributed from accretion is then given by
\begin{equation}
\rho_{\rm acc}(z)=\int_{\infty}^z\left(\frac{\rmd t}{\rmd z}\right){\rmd z}
   \int_{L_*(z)}^{\infty}\frac{1-\eta}{\eta}\frac{L_{\rm Bol}}{c^2}\Psi(L,z)\rmd L,
\end{equation}
where $L_*(z)$ is the limit luminosity due to survey sensitivity, $\Psi(L,z)$ is
the luminosity function, the bolometric luminosity is given by $L_{\rm Bol}=\calcb L_{\rm B}$, 
with $\calcb$  the correction factor relating $B-$band luminosity $L_{\rm B}$ to 
$L_{\rm Bol}$, $\eta$  the 
radiative efficiency and $c$ the light speed. This density includes {\em all} the black holes
that were brighter than $L_*(z)$.
The mass density of the active black holes at $z$ with $L>L_*$  is given by
\begin{equation}
\rho_{\rm qso}(z)=\int_{\mbh^{\rm C}}^{\infty}\Phi(\mbh,z)\mbh\rmd\mbh,
\end{equation}
where $\mbh^{\rm C}$ is the limit black hole mass due to the limit luminosity.
We apply the So\l tan's argument to the present sample at any redshift $z$ 
\begin{equation}
\rho_{\rm tot}(z)=\int_{\mbh^{\rm C}}^{\infty}{\cal N}(\mbh,z)\mbh\rmd\mbh=\rho_{\rm acc}(z).
\end{equation}
Though we do not know the distribution $\caln$, we know the mass density of all black holes
from So\l tan's argument. Finally, we have duty cycle $\bar{\delta}(z)$, 
\begin{equation}
{\bar \delta}(z)=\frac{\rho_{\rm qso}(z)}{\rho_{\rm acc}(z)}.
\end{equation}
If we know $\Psi(L,z)$ and $\Phi(\mbh,z)$, we can easily get 
the duty cycle of quasars at redshift $z$. We have to stress that Eq. (7) neither needs an initial 
condition nor the assumption of a constant Eddington ratio, something which must be specified for in
the alternative approach solving the continuity equation (e.g. $\bar{\delta}=1$ at $z=3$ in Marconi et al. 2004).

\section{Sample and black hole mass function}

\subsection{Spectrum Analysis}
The largest quasar sample, given by Richards et al. (2006) from the SDSS DR 3,
consists of 15,343 quasars from $z=0$ to 5, which is complete and homogeneous for apparent 
magnitude $i=15-19$. We only use those quasars (11,954) with $z\le 2.1$ in the present paper. 
We subtract the continuum and iron emission (based on the iron template derived from 
I Zw 1 spectra, which is kindly provided by R. J. McLure from private 
communication 2006). We then fit 
H$\beta$ and Mg {\sc ii} lines. For those with $z\le 0.7$, four components are used to model the
spectra: broad and narrow H$\beta$ plus narrow [O {\sc iii}] 4959\AA~ and [O {\sc iii}] 5007\AA.
For others, we use one broad and one narrow component to model the Mg {\sc ii} line.
Some objects are removed from the sample for one of three reasons; they have: 1) too poor spectra to fit 
due to low signal-noise ratio; 2) only narrow lines ($<2000~\kms$)
or 3) serious absorption at the Mg {\sc ii} line.
Finally we have 10,979 quasars available for estimating black hole masses
with $z\le 2.1$. The removed quasars reduce the completeness of the
sample given in Table 1. The last bin ($z_{10}$) is poor because the quality of quasar spectra
is not good enough to measure width of Mg {\sc ii}. A future paper will give a detailed description of 
estimating black hole masses and related issues (Chen et al. in preparation).

\subsection{Black Hole Mass Function}
 We apply the latest version of
the empirical relation of reverberation mapping (Kaspi et al. 2005; Vestergaard \& Peterson 2006)
to calculate the black hole mass in each quasar. 
For low redshift quasars with $z\le 0.7$, we use the full-width-half-maximum of 
H$\beta$ whereas we use Mg {\sc ii} for those with $0.7<z\le2.1$ (McLure \& Dunlop 2004). 
The scatter of the re-calibrated relation for the black hole masses is
less than 0.4 dex, which is much improved. Figure 1{\em a} shows the mass 
distribution of the present sample. We fit the mass distribution of the black holes via
least square method in a form of three power laws
\begin{equation}
{\cal F}(\mmbh)=f_0\mmbh^{\alpha_1}
\left(1+\frac{\mmbh}{m_{\bullet 1}}\right)^{-\alpha_2}
\left(1+\frac{\mmbh}{m_{\bullet 2}}\right)^{-\alpha_3},
\end{equation}
where $\mmbh$ is mass of black holes in unit of solar mass.
This expression has the limits ${\cal F}(\mmbh)\propto \mmbh^{\alpha_1}$ for $\mmbh\ll m_{\bullet 1}$,
${\cal F}(\mmbh)\propto \mmbh^{\alpha_1-\alpha_2}$ for $m_{\bullet 1}\ll\mmbh\ll m_{\bullet 2}$ and 
${\cal F}(\mmbh)\propto \mmbh^{\alpha_1-\alpha_2-\alpha_3}$ for $\mmbh\gg m_{\bullet 2}$. 
We obtain $f_0=(2.70\pm 0.35)\times 10^{-25}$, $m_{\bullet 1}=(4.14\pm 0.56)\times 10^7$, 
$m_{\bullet 2}=(2.50\pm 0.22)\times 10^9$, 
$\alpha_1=3.56\pm 0.28$, $\alpha_2=2.87\pm 0.25$ and $\alpha_3=2.71\pm 0.07$. A significant break appears at 
$\mbh=2.5\times 10^9\sunm$ and then a steeper mass spectrum ${\cal F}(\mmbh)\propto \mmbh^{-2.02}$
follows the break mass, which is consistent with the maximum mass from the SDSS DR1
(McLure \& Dunlop 2004).

{\footnotesize
\begin{center}{\sc Table 1  The mass function }\end{center}
\begin{center}
\begin{tabular}{ccrcccc} \hline\hline
$z$-bin   & $\Phi_* $     & $\beta_1$~~~&  $\beta_2$   &$\mbh^*$  & $\chi_j^2/d.o.f.$   &  $\cal{C}$ \\ 
          & $(10^{-6})$   &          &              & $(10^{9}\sunm)$ &     &  ($\%$)   \\ \hline
$z_1$     &        48.1   & $-$0.60      &    1.55      &   0.01    &    4.40/4         &    98  \\ 
$z_2$     &        4.74   &    0.32      &    2.48      &   0.16    &    7.90/9         &    95  \\
$z_3$     &        5.94   &    0.79      &    2.58      &   0.18    &    8.77/10        &    93  \\
$z_4$     &        0.79   &    0.18      &    2.89      &   0.58    &    9.89/10  	&    94 \\
$z_5$     &        0.44   &    0.48      &    3.36      &   1.25    &    8.25/10 	&    95 \\ 
$z_6$     &        0.27   &    0.15      &    3.21      &   1.81    &    8.16/9 	&    96 \\     
$z_7$     &        0.29   &    0.42      &    2.82      &   1.53    &    8.78/8 	&    97 \\         
$z_8$     &        0.33   &    0.68      &    2.86      &   1.55    &    8.58/8 	&    96 \\         
$z_9$     &        0.28   &    0.50      &    2.54      &   1.29    &    6.45/7 	&    89 \\
$z_{10}$  &        0.15   &    0.63      &    2.30      &   1.26    &    6.71/7  	&    67 \\ \hline
\end{tabular}
\parbox{3.1in}
{\baselineskip 9pt
The last column is the completeness at each redshift bin in our sample. The redshift bin $z_j$ is defined
by $z_j=(j-1/2)\Delta z$, where $j=1,..., 10$.
The fittings are not included one point in $z_1$, $z_4$, $z_6$ and $z_8-z_{10}-$bins, 
which significantly deviate from the trend of double power-laws.}
\end{center}
}

We get the MF by dividing our sample into 10 redshift bins 
with an interval $\Delta z=0.21$ and then into 20 black hole mass bins in
each redshift bin. Figure 1{\em b} shows the MF. 
We find that the function can be well fitted by double power laws in the following form
\begin{equation}
\Phi(\mbh,z)={\Phi_*}\left[\left(\frac{\mbh^*}{\mbh}\right)^{\beta_1}+
             \left(\frac{\mbh}{\mbh^*}\right)^{\beta_2}\right]^{-1},
\end{equation}	    
where $\beta_1$, $\beta_2$, $\Phi_*$ and $\mbh^{\ast}$ are constants. The peak mass is given by
$\mbh^{\rm peak}=\left(\beta_1/\beta_2\right)^{1/(\beta_1+\beta_2)} \mbh^*$.
We get the four parameters from the least $\chi-$square method via minimizing  
$\chi_j^2=\sum_i\left[\Phi(\mbh,z)-\Phi_{ij}\right]^2/\sigma_{ij}^2$, where 
$\Phi_{ij}$ is the MF derived from the sample in mass $M_{\bullet,i}-$bin and
redshift $z_j-$bin. Since the averaged error
bar in the luminosity function is about $\Delta \Psi/\Psi\sim 0.1$ (Richards et al. 2006), we take 
$\sigma_{ij}=0.1\Phi_{ij}$ for the MF. Table 1 lists all the parameters for each redshift bin.

\figurenum{1}
\centerline{\psfig{figure=f1.ps,angle=270,width=8.0cm}}
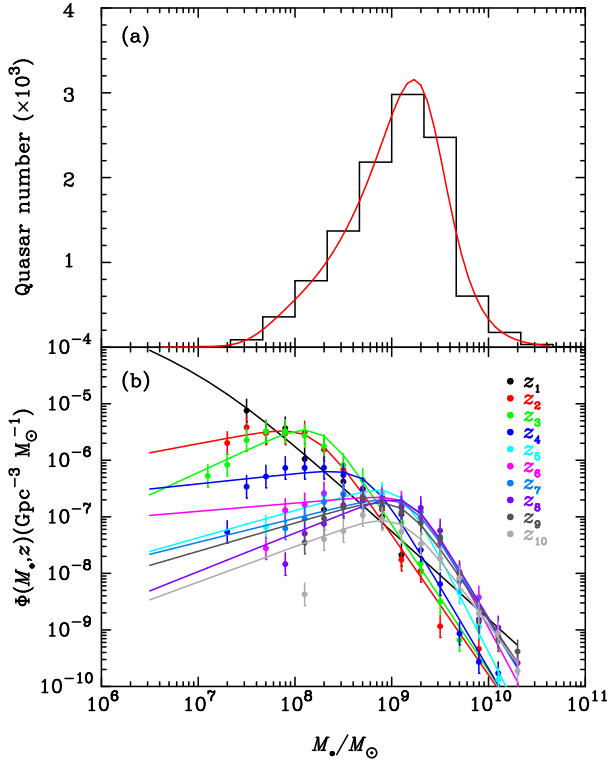
\figcaption{{\em a} shows the mass distribution of the black holes in our sample. 
The red line is the least square fit. The black hole mass function 
$\Phi(\mbh,z)$ at each redshift bin is shown in {\em b}. The lines represent the best fits given in Tab 1.}
\label{fig1}
\vglue 0.3cm

The MF shows an increasing peak mass toward high redshifts. The mass break at the low mass 
side is not real. It is caused by the survey flux limit. We have to point out that the MF 
is {\em not} complete for all black holes, but it is complete for our flux-limited criterion.

\section{Results}
The $B-$band luminosity is converted from the absolute magnitude $M_i(z=2)$ via a relation
of $M_B=M_i(z=2)+0.804$ (Richards et al. 2006).
It has been found that the factor $\calcb$ is not a function of redshift (
Steffen et al. 2006). A more 
elaborate treatment gives $\calcb=6-7$ for quasars (Marconi et al. 2004). 
We use $\calcb=6.5$ throughout the redshift range in this 
paper. The luminosity function is taken from Richards et al. (2006).
We take $\mbh^{\rm C}$ from the minimum mass of the black holes in each redshift bin to calculate 
their mass density, $\rho_{\rm qso}$(Eq. 5). This corresponds to the survey limit and is consistent with 
the accretion density $\rho_{\rm acc}$ (Eq. 4).
The radiative efficiency $\eta\ge 0.1$ is reached by Yu \& Tremaine (2002), $\eta\ge 0.15$
by Elvis et al. (2002) and $\eta=0.15$ by Gammie et al. (2004), based on the So\l tan's argument,
the X-ray background and numerical simulations, respectively. Numerical calculations indicate that black 
holes are rotating with their maximum spin all the time during their evolution if accretion 
is included (Volonteri et al 2005), which is supported by studies of SDSS quasars (Wang et al. 2006);
then $\eta\approx 0.3$ without significant evolution. We thus calculate 
the duty cycle for two different radiative efficiencies $\eta=0.1$ and $0.3$. 

\figurenum{2}
\centerline{\psfig{figure=f2.ps,angle=270,width=8.0cm}}
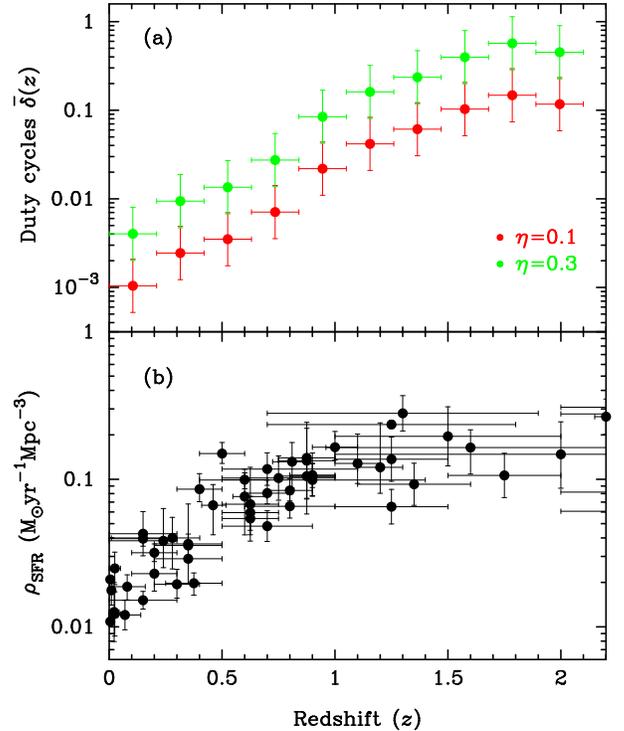
\figcaption{\footnotesize {\em a, b} show the duty cycle of quasars as a function of redshift in our 
sample and the history of the SFR density, respectively.  The SFR density
is taken from P\'er\'ez-Gonz\'alez et al. (2005).
}
\label{fig2}
\vglue 0.3cm

Results are shown in Figure 2.  First, quasars have duty cycle of $\bar{\delta}(z)=10^{-3}\sim 1$
as shown in Fig 2{\em a}, indicating that black holes are undergoing active and dormant phases, namely 
episodic activity. The value of $\bar{\delta}(z)\rightarrow 1$ at $z\sim 2$ agrees with the assumption 
of $\delta=1$ at $z=3$ in Marconi et al. (2004).
Second, the duty cycle is rapidly evolving from $z\approx 2$ to the local Universe. We find 
$\bar{\delta}(z)\propto z^{\gamma}$, where $\gamma\sim 2.5$ until $z=1.5$, above which it tends to flatten. 
Third, as shown in Fig. 2{\em a} and 2{\em b}, the duty cycle connects quite naturally with the history 
of the SFR density as a consequence of co-evolution of galaxies and black holes. 
Massive, gas-rich mergers account not only for most of the star formation at 
$z\approx 2-3$, but they are probably also responsible for triggering major episodes of 
black hole activity (Di Matteo et al. 2005). Figures 2{\em a} and 2{\em b} show that the duty cycle follows 
the star formation history, implying evidence for
intrinsic star-formation-triggered black hole activity. The higher the SFR density,
the higher the triggering frequencies of the black hole activity. Last, we define
${\cal R}=\Delta t_{\rm act}/\Delta t_{\rm dor}$, then have
${\cal R}=\bar{\delta}(z)/\left[1-\bar{\delta}(z)\right]$.
Episodic activity of the black holes can be described by this parameter.
When $\bar{\delta}(z)\rightarrow 1$,
black holes have ${\cal R}\gg 1$, namely, the dormant black holes are frequently
triggered at $z\sim 2$ by star formation. 
At that time, quasars look like long-lived phenomena because of ${\cal R}\gg 1$.

\section{Discussion and Summary}
The evolution of quasars is jointly controlled by the triggering mechanism and accretion.
The duty cycle is a key parameter to unveil the evolution of quasars. The results of the present paper show 
a very strong cosmological evolution of quasar's duty cycle. 
The triggering history represented by $\bar{\delta}(z)$ is quite similar to the
evolution of cosmic SFR density. This indicates that star formation may be the direct 
mechanism to trigger the activity of the black holes.

The duty cycle can be roughly justified from the galaxy luminosity function. 
According to the luminosity function of galaxies at $1.8\le z\le 2.0$ (Dahlen et al. 2005), the galaxy
number density is $n_{\rm G}\approx 826$Gpc$^{-3}$ for galaxies brighter than $R-$band magnitude 
$M_{\rm R}=-24$. This corresponds to the number density of galaxies with black hole mass larger than
$10^9\sunm$ converted from $\log \left(M_{\rm BH}/\sunm\right)=-0.5M_{\rm R}-3$ (McLure \& Dunlop 2001).
Number density of quasars brighter than $M_i=-28$ (corresponding to a black hole with mass $>10^9\sunm$
if it is accreting at the Eddington limit) is $n_{\rm Q}\approx 175$ Gpc$^{-3}$ based 
on the quasar luminosity function (Richards et al. 2006). We thus estimate a duty cycle of $\sim 0.18$, which is
roughly consistent with the present results (see Figure 2a).

The SFR density rises with redshift out to $z=1.5$ and appears to be roughly flat between $z\approx 1.5$ and 
$z\approx 3.0$. This tendency
could be explained by strong feedback from activity of black holes to their host galaxies 
(Silk 2005; Di Matteo et al. 2005; Croton et al. 2006).  With a balance between star formation and 
feedback in $z\sim 1.5-3.0$, the violent star formation is then suppressed.
However, the SFR density is going to decrease with time due to
a shortage of gas and the BH duty cycle follows this trend. To further confirm this, future work
will focus on the dependence of the duty cycle on the black hole masses. It could show the
feedback-dependence on the BH growth itself. Additionally the total accretion time
(net lifetime) of black holes will be then obtained.

\acknowledgements{The anonymous referee is greatly acknowledged for pointing out a mathematical error in
the paper. The authors are grateful to M. Z. Kong, R. Wang and R. J. McLure for help in SDSS data 
analysis, and L. C. Ho, S. N. Zhang, J. X. Wang and X. Y. Xia for useful discussions.  This research is 
supported by the Natural Science Foundation of China through NSFC-10325313, 10233030 and 10521001. }

\end{document}